\documentclass[12pt]{article}
\usepackage{epsf}
\def\be{\begin{equation}}
\def\ee{\end{equation}}
\def\bea{\begin{eqnarray}}
\def\eea{\end{eqnarray}}

\begin{document}
\begin{titlepage}
\begin{center}
{\Large \bf William I. Fine Theoretical Physics Institute \\
University of Minnesota \\}
\end{center}
\vspace{0.2in}
\begin{flushright}
FTPI-MINN-15/44 \\
UMN-TH-3507/15 \\
October 2015 \\
\end{flushright}
\vspace{0.3in}
\begin{center}
{\Large \bf Remarks on measurement of the decay $\Xi_b^- \to \Lambda_b \pi^-$ 
\\}
\vspace{0.2in}
{\bf  M.B. Voloshin  \\ }
William I. Fine Theoretical Physics Institute, University of
Minnesota,\\ Minneapolis, MN 55455, USA \\
School of Physics and Astronomy, University of Minnesota, Minneapolis, MN 55455, USA \\ and \\
Institute of Theoretical and Experimental Physics, Moscow, 117218, Russia
\\[0.2in]

\end{center}

\vspace{0.2in}

\begin{abstract}
The interpretation of the recent observation by LHCb of a heavy-flavor-conserving and strangeness-changing decay $\Xi_b^- \to \Lambda_b \pi^-$ in terms of the branching fraction for the decay suffers from the uncertainty in the yield of the $\Xi_b^-$ hyperons relative to that of the $\Lambda_b$. It is pointed out here that this relative yield can be determined with a significantly reduced uncertainty by measuring within the same experimental conditions the decays related by the flavor $SU(3)$ symmetry and/or the heavy quark symmetry, such as $\Xi_b \to J/\psi \, \Xi$, $\Lambda_b \to J/\psi \, \Lambda$, or $\Xi_b^- \to \Xi_c^0 \ell \nu$, $\Lambda_b \to \Lambda_c \ell \nu$.

\end{abstract}
\end{titlepage}

The LHCb collaboration has recently reported~\cite{lhcb} experimental evidence for the strange\-ness-changing weak decay $\Xi_b^- \to \Lambda_b \pi^-$. Unlike the dominant decays of the $b$ hadrons due to the disintegration of the heavy quark,  
the  $\Xi_b$ hyperon decays of this type proceed through the weak decay of the strange quark. Such processes, similar to the weak decays of ordinary strange hyperons, were discussed some time ago~\cite{ccllyy,sk,mv00} and were recently revisited in the literature~\cite{lv,fm}. These sub-dominant decays of the heavy hyperons are of an interest due to their relation to the dynamics of the quarks in baryons~\cite{mv00,lv}, and barring the possibility of unusually strong diquark correlations~\cite{vzs,djs} in the baryons, their rate is expected~\cite{mv00,fm} in the same ballpark as that for ordinary hyperons, which amounts to being of the order of 1\% of the total decay rate of the $\Xi_b$ heavy hyperons. 

The LHCb experiment reported a result for the observed rate of the decay  $\Xi_b^- \to \Lambda_b \pi^-$ relative to the observed yield of of the $\Lambda_b$ hyperons:
$(f_{\Xi_b^-}/f_{\Lambda_b}) \, {\cal B} (\Xi_b^- \to \Lambda_b \pi^-) = (5.7 \pm 1.8^{+0.8}_{-0.9}) \times 10^{-4}$. Naturally, the interpretation of this result in terms of the branching fraction alone requires knowledge of the ratio of the yield of $\Xi_b^-$ and $\Lambda_b$: $R=f_{\Xi_b^-}/f_{\Lambda_b}$. In the LHCb paper~\cite{lhcb} a broad range $R \approx 0.1 - 0.3$ is assumed ``based on measured production rates of other strange particles relative to their non-strange counterparts'' resulting in the estimate for the branching fraction ${\cal B} (\Xi_b^- \to \Lambda_b \pi^-) \approx (0.2 - 0.6)\%$. Clearly, this estimate leaves a better quantitative accuracy to be desired. The purpose of the present note is to point out that the ratio $R$ can be determined by measuring the relative yield of decays of the $\Xi_b^-$ hyperons and similar decays of the $\Lambda_b$ where the ratio of the absolute decay rates can be evaluated with some theoretical certainty. In other words, the ratio of the event count for the processes $\Xi_b \to X_\Xi$ and $\Lambda_b \to X_\Lambda$ with the corresponding final states $X_\Xi$ and $X_\Lambda$ after correcting for the specific experimental efficiencies is given by
\be
{N(X_\Xi) \over N(X_\Lambda)}=R \, {{\cal B}(\Xi_b^- \to X_\Xi) \over  {\cal B}(\Lambda_b^- \to X_\Lambda)}  =  R \, {\tau(\Xi_b^-) \over \tau(\Lambda_b)} \, {\Gamma(\Xi_b^- \to X_\Xi) \over  \Gamma(\Lambda_b^- \to X_\Lambda)}~.
\label{nr}
\ee
Since the ratio of the lifetimes is now known with a reasonable accuracy~\cite{lhcbt}: $\tau(\Xi_b^-)/\tau(\Lambda_b) = 1.089 \pm 0.028$, the ratio $R$ can be determined provided that the ratio of the absolute decay rates $\Gamma(\Xi_b^- \to X_\Xi) /  \Gamma(\Lambda_b^- \to X_\Lambda)$ can be evaluated theoretically.

One type of such similar decay of the $\Xi_b^-$ and $\Lambda_b$ is generated by the underlying $b \to c \bar c s$ weak interaction process: $\Xi_b^- \to J/\psi \, \Xi^-$ and $\Lambda_b \to J/\psi \, \Lambda$. The absolute rates of these decays are simply related by the flavor $SU(3)$ symmetry. Indeed, the final states in these decays are in an $SU(3)$ octet, while the initial heavy hyperons belong to an $SU(3)$ antitriplet, so that there is only one $SU(3)$ invariant amplitude for the decay due to the $b \to c \bar c s$ weak interaction, which is an $SU(3)$ triplet. One thus readily finds~\footnote{It is clear that the same relation holds for a final state with any state of $c \bar c$ charmonium in  place of $J/\psi$.}
\be
\Gamma(\Lambda_b \to J/\psi \, \Lambda) = {2 \over 3} \, \Gamma (\Xi_b^- \to J/\psi \, \Xi^-)~.
\label{jpr}
\ee
The uncertainty in this relation is the usual one for the flavor $SU(3)$ predictions that can receive corrections of the first order in the symmetry breaking. Conventionally this uncertainty is estimated as 30\%, although it can be smaller for specific relations.

It can be noted that for these decay processes the ratio equivalent to that in Eq.(\ref{nr}) is available, albeit for the conditions of the CDF and D0 experiments at Tevatron. Namely, the Particle Data Group~\cite{pdg} quotes ${\cal B}(\Xi_b^- \to J/\psi \, \Xi^-) \, f(b \to \Xi_b^-) = (1.02^{+0.26}_{-0.25}) \times 10^{-5}$  and ${\cal B}(\Lambda_d \to J/\psi \, \Lambda) \, f(b \to \Lambda_b) = (5.8 \pm 0.8) \times 10^{-5}$ as the average of the CDF~\cite{cdf1,cdf2} and D0~\cite{d01,d02} data. Using these average values and the relation (\ref{jpr}) one finds from Eq.(\ref{nr}) that in the kinematical conditions of the Tevatron experiments the ratio $R$ is given by
\be
R_{\rm Tev} = 0.11 \pm 0.03~,
\label{rd}
\ee
with theoretical uncertainty being comparable to the indicated experimental one. One can readily notice that this estimate is near the lower limit of the assumed in Ref.\cite{lhcb} range of the values of $R$. It is certainly possible that at the LHC energy and with the kinematical cuts inherent in the LHCb experiment the value of $R$ is different from that in the CDF and D0 experiments. It thus would be quite helpful if the ratio of the yield of events $(\Xi_b^- \to J/\psi \, \Xi^-)/(\Lambda_b \to J/\psi \, \Lambda)$ is measured within the specific kinematical conditions of the LHCb experiment.

The theoretical uncertainty due to the violation of the flavor $SU(3)$ symmetry in the relation (\ref{jpr}) can be substantially reduced if instead one uses relations for pairs of similar semileptonic decays of the $\Xi_b$ and $\Lambda_b$ hyperons. For such decays one can consider the inclusive final states $\Xi_b \to X_{cs} \ell \nu$ and $\Lambda_b \to X_c \ell \nu$ with $X_c$ ($X_{cs}$) being a final state with charm (and strangeness), or the semi-inclusive $\Xi_b \to \Xi_c \ell \nu + anything$ and $\Lambda_b \to \Lambda_c \ell \nu+ anything$, or the exclusive ones $\Xi_b \to \Xi_c \ell \nu$ and $\Lambda_b \to \Lambda_c \ell \nu$. The best theoretical accuracy is guaranteed for the inclusive channel. According to the heavy quark expansion (HQE)~\cite{sv} (a recent review can be found in Ref.~\cite{lenz}), the rate of the inclusive decay with charm in the final state should be the same for $\Xi_b$ and $\Lambda_b$ within an accuracy of order 1\% or better. In practice the semi-inclusive channels, i.e. explicitly containing the corresponding charmed hyperon $\Xi_c$ or $\Lambda_c$ should be almost as good as the inclusive ones, since it is known~\cite{pdg} that the decay $\Lambda_b \to  
\Lambda_c \ell \nu+ anything$ practically saturates the semileptonic branching ratio for $\Lambda_b$ and there is every reason to expect that a similar behavior should be true for the semileptonic decays of $\Xi_b$. Finally, the exclusive semileptonic decay modes of each of the $b$ hyperons are described  by the heavy quark symmetry~\cite{rv,iw}. In the symmetry limit these decays should have the same rate:
\be
\Gamma(\Xi_b \to \Xi_c \ell \nu) = \Gamma(\Lambda_b \to \Lambda_c \ell \nu)~,
\label{slr}
\ee
so that any corrections to this relation require a combined effect of deviation from the heavy quark symmetry and of breaking of the $SU(3)$ symmetry between the light quarks. The most theoretically stringent relation arises~\cite{rv} for the differential rates of the decays in Eq.(\ref{slr}) near the maximum invariant mass of the lepton pair, i.e. near the point of zero recoil for the charmed hyperon. At this point the relative corrections are quadratic in the deviation from the heavy quark symmetry, and e.g. for the $B$ mesons are known~\cite{bsu} not to exceed 10\%. Alowing, conservatively,  for an additional 30\% difference of these corrections between the decays of $\Xi_b$ and $\Lambda_b$ due to the $SU(3)$ breaking one can count on the accuracy of just a few percent for the relation between the semileptonic decay spectra near zero recoil. 

In conclusion. It is quite clear that the current uncertainty in the measurement of ${\cal B}(\Xi_b^- \to \Lambda_b \, \pi^-)$ amounting to a factor of three due to the yet undetermined ratio $R$ of the yield of $\Xi_b^-$ and $\Lambda_b$ hyperons should and can be greatly reduced.  The way to such reduction, discussed here, is a measurement of similar decay channels for the two hyperons, whose absolute rates are related by the known symmetries. For one such pair of decays, $\Xi_b^- \to J/\psi \, \Xi^-$ and $\Lambda_b \to J/\psi \, \Lambda$, the relation (\ref{jpr}) between the absolute rates relies on the $SU(3)$ symmetry, whose `nominal' accuracy is about 30\% (which is still better than the current factor of three uncertainty). A more theoretically accurate relation between the absolute decay rates can be argued for the semileptonic decays rates of the two hyperons [as in Eq.(\ref{slr})] where the relation is protected by the heavy quark symmetry in addition to the flavor $SU(3)$.

I thank S.~Blusk and A.Bondar for very helpful discussions  and for the encouragement to publicize this note in spite of its simplicity. This work is supported in part by U.S. Department of Energy Grant No.\ DE-SC0011842.

\end{document}